# Collective magnetic response of $CeO_2$ nanoparticles

Michael Coey, Karl Ackland, Munuswamy Venkatesan and Siddhartha Sen,

School of Physics and CRANN, Trinity College, Dublin 2, Ireland

**The magnetism of nanoparticles and thin films of wide-bandgap oxides that include no magnetic cations is an unsolved puzzle[1]. Progress has been hampered both by the irreproducibility of much of the experimental data, and the lack of any generally-accepted theoretical explanation. The characteristic signature is a virtually anhysteretic, temperature-independent magnetization curve which saturates in an applied field that is several orders of magnitude greater than the magnetization. It appears as if a tiny volume fraction, <~ 0.1%, of the samples is magnetic and that the energy scale of the problem is unusually high for spin magnetism. Here we investigate the effect of dispersing 4 nm $CeO_2$ nanoparticles with powders of $\gamma Al_2O_3$, sugar or latex microspheres. The saturation magnetization, $M_s \approx 60$ Am$^{-1}$ for compact samples, is maximized by 1 wt% lanthanum doping. Dispersing the $CeO_2$ nanopowder reduces its magnetic moment by up to an order of magnitude. There is a characteristic length scale of order 100 nm for the magnetism to appear in $CeO_2$ nanoparticle clusters. The phenomenon is explained in terms of a giant orbital paramagnetism that appears in coherent mesoscopic domains due to resonant interaction with zero-point fluctuations of the vacuum electromagnetic field. The theory explains the observed temperature-independent magnetization curve and its doping and dispersion dependence, based on a length scale of 300 nm that corresponds to the wavelength of a maximum in the UV absorption spectrum of the magnetic $CeO_2$ nanoparticles. The coherent domains occupy roughly ten percent of the sample volume.**

Corresponding author: jcoey@tcd.ie

There is a class of thin films and nanoparticles of oxides that exhibit ferromagnetic-like magnetization curves, although the oxides lack the concentration of ions with unpaired *d* or *f* electron spins required to generate the exchange interactions needed for high-temperature ferromagnetism[1]. This is forcing a re-evaluation of the meaning of magnetic saturation in systems that exhibit very little hysteresis. Research in this area has been plagued by a shortage of reproducible experimental data, so there is need for an easily-synthesised 'fruitfly' system that reliably exhibits stable anomalous magnetism for which no extraneous explanation is possible. The much-studied dilute Co-doped ZnO thin films[2] were problematic because metallic cobalt nanoparticles, difficult to detect in ZnO films[1-4], are ferromagnetic with a high Curie temperature.

The reports of magnetism in these oxide systems have shown that a 3*d* dopant is unnecessary, and even when 3*d* ions are present they do not necessarily order magnetically[5-8]. The magnetism is somehow related to defects in the oxide; candidates include cation[9] or oxygen[10] vacancies (F-centres). Sunderasen *et al*[11] have suggested that weak room-temperature magnetism could actually be a general feature of oxide nanoparticles. A significant observation was that the magnetism appearing in undoped 10 nm ZnO nanoparticles depends on how they are capped with different molecules, which alter the electronic structure of the surface[12]

A promising candidate system is $CeO_2$, where nanoparticles produced in different laboratories often exhibit weak 'ferromagnetic-like' behavior at room temperature. A selection of data is presented in Table 1. Although values of saturation magnetization $M_s$ are very small and vary widely, the saturation field $H_0$ obtained by extrapolating the initial susceptibility to saturation (See Figure 1d) is roughly 1000 times greater and lies in a narrower range of 40 - 120 kA m$^{-1}$. The ratio $\mathcal{N}_{eff}M_s/H_0$ with an effective demagnetizing factor $\mathcal{N}_{eff} \approx 0.3$ is a measure of the magnetic volume fraction $f$ in a ferromagnetic system where the approach to saturation is governed by dipolar interactions[1]. $f$ is of order 0.1 %.

We synthesized many small batches ( ~ 4 mg) of nanocrystalline $CeO_2$ powder using as precursor either high purity (99.999 %) $Ce(NO_3)_36H_2O$, or reagent grade (99 %, with La as the main impurity. The nanoparticles are well crystallized, and only $2r$ = 4 nm in diameter (Figure 1b). Magnetization curves of nanoparticles produced from the two precursors are compared in Figure 2a. The pure sample shows a practically linear paramagnetic response, but the impure sample exhibits a superposed ferromagnetic-like curve with no evidence of hysteresis. The averaged specific magnetization $\sigma_s$ for 16 samples is 0.011 ± 0.006 A $m^2$ $kg^{-1}$, corresponding to an average moment per Ce ion of 3 $10^{-4}$ $\mu_B$ and an average saturation magnetization $M_s$ = 84 $Am^{-1}$. These numbers can be misleading. There is evidence that the moment is associated with defects[13-15] or with the nanoparticle surface[12,16] rather than cerium ions or the volume, so it is better to think of a few tenths of a Bohr magneton as the average moment per particle. The content of Fe, Co and Ni impurities in the 99% $CeO_2$ nanopaticles – 10 ppm all told – is too little to account for the observed moments, since each nanoparticle contains approximately 900 cerium atoms.

In fact it is the lanthanum doping that is responsible for the moment. On doping the pure cerium nitrate precursor with pure lanthanum nitrate, there is a sharp maximum in magnetization for $Ce_{1-x}La_xO_2$ at $x$ = 1.0 % and the moment has almost disappeared at x = 10% (Figure 2b). A similar decline has been observed for Pr doping[14]. The moments are fairly stable in time, decaying by about a tenth over the course of a year. Moreover, the magnetization curves between 4 K and 380 K superpose after correcting for the high-field slope, Figure 2c, with little sign of hysteresis at any temperature. The insensitivity of the magnetization to thermal excitations, at least up to 380 K, is evidence that an unusually large energy scale > 0.1 eV must be involved.

Like $Ce^{4+}$, $La^{3+}$ has a $4f^0$ configuration, so La substitution in stoichiometric $CeO_2$ would normally create holes at the top of the oxygen $2p$ band. However, $CeO_2$ is a catalyst reknown for its oxygen vacancies, and the addition of La increases the quantities of both oxygen vacancies and peroxide ions that occur naturally at the $CeO_2$ surface.[17] The content of localized $Ce^{3+}$ ions estimated from the Curie-law variation of the susceptibility (Figure 2b) is only 0.4 %. If there are any other $Ce^{3+}$ electrons, they are delocalized at the bottom of the $4f$ band as suggested in Figure 1c). The $2p$-$4f$ gap for stoichiometric $CeO_2$ is 3.2 eV[15], and the $2p$-$5d/6s$ gap is 6 – 8 eV[18]. Electrons will tend to segregate to the nanoparticle surface, which is conducting for oxygen-deficient $CeO_2$[19.20]. It should be emphasized that $Ce^{3+}$compounds or intermetallics rarely order magnetically above 15 K (1.3 meV), and the maximum reported value is 125 K [21].

In a series of experiments where the $CeO_2$ nanopowder was progressively diluted with powders of different particle size, we used a 15 nm $\gamma Al_2O_3$ nanopowder, finely-ground icing sugar with an average particle size ~ 1 μm and latex microspheres 10 microns in diameter. The surprising effect of dispersion with $\gamma Al_2O_3$ is shown in Figure 3a. When a 4 mg sample of $CeO_2$ is diluted with six times its own volume of diamagnetic $\gamma Al_2O_3$, the magnetic moment collapses to just 6% of the original value (Figure 3) The effect of dilution is to separate clumps of $CeO_2$ nanoparticles ≤ 100 nm in size. Dilution with finely-ground sugar has a similar, if less dramatic effect. The moment there falls by 50% on diluting with 30 times the volume of sugar, which is less effective than $\gamma Al_2O_3$ at dispersing the $CeO_2$ particles, but has the advantage is that some of the $CeO_2$ can be recovered by dissolving the sugar in water. The specific magnetization for the recovered $CeO_2$ is double that at the outset. Large, 10 μm latex microspheres are least effective; the moment is reduced by 15 % for a 20-fold volume dilution. The aggregates of $CeO_2$ coexisting with the microspheres are about 500 nm in size, and sometimes envelop them (Fig. 3d).

Together, these experiments establish that the magnetism of $CeO_2$ depends critically on the *mesoscale disposition of the nanoparticles,* as well as doping, which is probably why there is so much variability in the data of Table 1. We conclude that there is a characteristic length scale for the appearance of magnetism, which is of order 100 nm. Previously, Radovanovich and Gamelin found that the moment of 6 nm nanoparticles of ZnO doped with 0.93% Ni only appeared in reaction-limited aggregates about 400 nm in size[22], while Sundaresan and Rao reported that the moment in 7 nm $CeO_2$ powder was modified upon sintering[13]. The magnetism is not simply an intrinsic property governed by atomic-scale defects within the particles. The extent and topology of the surface of contiguous particles is a critical factor.

Until now, a plausible model for the high-temperature magnetism of $CeO_2$ nanoparticles has been Stoner ferromagnetism with a spin-split band associated with conducting surface states[23]. Furthermore, if the band is half-metallic, spin-wave excitations are suppressed, and a high Curie temperature could be envisaged[24]. The problem is to understand how, when we break up the $CeO_2$ nanoparticle sample into 100 nm clumps, we can lose the magnetism. Stoner splitting of a 4*f* band or a defect band, of order a few tenths of an electron volt, will not change appreciably when the sample is divided up. The closest analogy in the conventional paradigm is the stabilization of magnetic order in clusters of superparamagnetic nanoparticles by dipole-dipole interactions[25]. The magnetite particle chains in magnetotactic bacteria are a good example. Contiguous particles with a magnetization of order 0.5 MA $m^{-1}$ and a moment of order 1000 $\mu_B$ can provide a dipole interaction energy that exceeds room temperature. The average moment of a $CeO_2$ nanoparticle is three or four orders of magnitude too small.

It appears that a radically new approach is required. We propose that the magnetic saturation is not related to collective spin ferromagnetism but to *giant orbital paramagnetism*[26] associated with the collective response of electrons in coherent domains to an applied magnetic field. Our starting point is a new model[27], which showed that when zero-point fluctuations of the vacuum electromagnetic field (EMF) interact with an ensemble of two-level atoms, it is possible for coherent mesoscopic domains to emerge. This can take place at room temperature in quasi-two-dimensional systems, with a large surface/volume ratio. No resonant cavity is required. The size of the coherent domains is determined by the wavelength corresponding to an electronic excitation $\hbar\omega$ between the ground state and the excited state of the two-level atoms. The excited state lies at an energy $\varepsilon$ below the ionization threshold, as illustrated in Fig 4a). The interaction of the $N$ electrons in a coherent domain of size $\lambda \approx 2\pi c/\omega$ with the vacuum field leads to stabilization of the ground state and destabilization of the excited state each by an energy $G^2\hbar\omega$ where $G$ is calculated to be $\approx 0.1$ [27]. The model is parameterized in terms of $N$, $\omega$ and $\varepsilon$, and the stability condition is $k_BT < G^2\hbar\omega < \varepsilon$.

In the coherent ground state, the electrons have a common oscillation frequency $G^2\omega$ and a corresponding moment (see Supplementary Information)

$$\mu_c = [(2l + 3)(2l + 4)/8](G^2\hbar\omega/2\varepsilon)\mu_B \qquad (1)$$

that is set by the size of the orbit, where $l$ is the orbital quantum number of the electronic ground state and $\mu_B$ is the Bohr magneton. In the presence of the time-varying vacuum electromagnetic field, the effect of a static magnetic field on the coherent domain is to produce a modified coherent ground state, inducing a paramagnetic orbital moment in the domain that is nonlinear in $B$ and proportional to $\sin 2\alpha_m$, where $\alpha_m$ is a mixing angle (See Supplementary Information). The magnetization curve has the form

$$M = M_s x/(1 + x^2)^{1/2} \qquad (2)$$

where $x = CB \approx GN\mu_c B/\hbar\omega$. This function differs only slightly from the empirical $M = M_s \tanh y$ function often used to fit magnetization curves[23], but it follows directly from theory. Fits of Eq. 2 to the curves in Figure 2c at 4 K, 295 K and 380 K give very similar fit parameters $C = 9.4 \pm 0.7\ T^{-1}$ and $M_s = 58 \pm 1\ A\ m^{-1}$.

The length scale in the problem is set by the characteristic excitation frequency ω of $CeO_2$, which is resonant with the zero-point vacuum fluctuations. The corresponding wavelength is $\lambda = 2\pi c/\omega$, and the volume of the coherent domain is $v_c \approx (\pi/6)\lambda^3$. In Fig 1d) there is a prominent absorption at $\lambda = 300$ nm in the UV spectrum of the magnetic nanoparticles. The corresponding frequency of the electronic transition is $\omega = 6.3\ 10^{15} s^{-1}$, and the photon energy is $\hbar\omega = 4.1$ eV. No real photons of this energy are emitted or absorbed according to the theory[27]; it is the zero point energy that, due to its time-dependence, can mix states different in energy by $\hbar\omega$ to produce a modulated collective response frequency for all $N$ electrons in a coherent domain.

By fitting the magnetization curve to Eq 2, we can deduce the volume fraction $f_c$ of the sample that is composed of coherent domains, and estimate their magnetic moment $N\mu_c$. Dividing the saturation magnetization $M_s \approx f_c GN\mu_c/2v_c$, by $C \approx$ $GN\mu_c/\hbar\omega$, we obtain $M_s/C = f_c\hbar\omega/2v_c$. With the experimental value of $M_s/C$ and $G \approx 0.1$, we find $f_c = 28$ % and the coherent domain moment $N\mu_c = 6.6\ 10^6\mu_B$. Identifying $N$ with the number of La dopant atoms in a coherent domain (2.4 $10^6$), the coherent moment per dopant $\mu_c = 2.8\ \mu_B$.

The orbital moment expected from Eq 1 depends on $G$, the orbital quantum number of the ground state and the ionization energy ε of the excited state. Taking $G \approx 0.1$ and $l = 3$, identifying the transition that becomes very prominent in the magnetic nanoparticles (Figure 1d) as a $4f - 5d$ transition, the ionization energy $\varepsilon \approx 0.1$ eV. The values of $G$ and $N$ are in accord with the values anticipated for quasi-two dimensional coherent domains in the model[27] ($G \approx 0.1$ and $N \approx 10^6$ for $\hbar\omega = 4.1$ eV). The model of Ref 27 was simplified, and took no account of the spin of the electrons. The influence of spin-orbit coupling and taking account of Fermi-Dirac statistics in these dilute electronic systems will not modify the semi-quantitative agreement between the theory and our experiments. Generally, it will be possible to estimate the size λ of the coherent domains in such systems by fitting the magnetization curve to Eq 2 in order to determine $C/M_s$ and then using the following expression that follows from $\hbar\omega = 2\pi\hbar c/\lambda$, $M_s/C = f_c\hbar\omega/2v_c$ and $v_c = \pi\lambda^3/6$

$$\lambda = [(C/M_s)(6\hbar c f_c)]^{1/4} \sim [(C/M_s)\hbar c]^{1/4} \qquad (3)$$

In summary, giant orbital moments related to zero-point fluctuations of the vacuum electromagnetic field can resolve a long-standing problem in magnetism. The theory accounts for the temperature-independence of the magnetization curve, and the characteristic length scale of order 100 nm required for the appearance of the magnetically-induced orbital currents in coherent domains. The saturation magnetization $M_s$ and the parameter $C$ in Eq 2, which are easily obtained by fitting the measured magnetization curve, determine the size of the coherent domains via Eq 3 and their giant orbital moments, assuming $G \approx 0.1$. The data on all our magnetic samples and almost all the other data in Table 1 are consistent with $\lambda \approx 300$ nm, and coherent volume fractions of 1 – 80 %.

Giant orbital paramagnetism is a new observable consequence of zero-point electromagnetic energy — it is the first such magnetic effect. It occurs in mesoscopic quasi-2D matter where the active sites are dilute and the effects of Fermi-Dirac statistics can be neglected. Spin-orbit interaction is expected to stabilize the coherent state. It is anticipated that the present study will lead further investigations of measurable consequences of resonant zero-point fluctuations not only in magnetic systems such as gold nanoparticles[28], but in other areas of condensed matter, whether physics, chemistry or biology. Some candidate systems are nanobubbles, the water/cell interface and concentrated ionic solutions.

*Methods.* The $CeO_2$ nanoparticles were synthesised by homogeneous precipitation[23] from 10 mM $Ce(NO_3)_3 6H_2O$ solutions by dropwise addition of 0.1 M NaOH (99.99% purity). 0.45 mL (1/10 the volume of $Ce(NO_3)_3 6H_2O$ solution) of 0.5 M Polyethylene glycol (PEG) with molecular weight 1500 is added to help separate the nanoparticles during formation, which are well crystallized, but only about 4 nm in size (Figure 1). Magnetization was measured on 4 mg samples of $CeO_2$ nanoparticles using a 5 T Quantum Design SQUID magnetometer. Powders were contained in gelcap sample holders, which produce a linear diamagnetic response, mounted in a plastic straw. All isothermal magnetization curves, but not the thermal scans, have been corrected for the linear response.

Acknowledgement. This work was supported by Science Foundation Ireland as part of the NISE project, contract 10/IN1.I3006

*Contributions:* JMDC conceived the experiment and wrote the paper, KA and MV carried out the experiments and reduced the data, SS developed the theory, JMDC and SS and analyzed the results.

*References.*

Table 1. A selection of magnetization data reported for $CeO_2$ nanoparticles, together with the extrapolated saturation field $H_0$ and the magnetic volume fraction $f$.

| Average radius $r_0$ (nm) | $M_s$ ($Am^{-1}$) | $H_0$ ($kAm^{-1}$) | $f$ * ($10^{-6}$) | Surface treatment | Reference |
|---|---|---|---|---|---|
| 3.5 | 7 | 60 | 39 | - | a |
| 7.5 | 11 | 40 | 92 | - | a |
| 5×1 | 550 | 80 | 2290 | PEG | b |
| 3 | 40 | 80 | 168 | Oleic acid | c |
| 3.5 | 1.5 | 120 | 4 | Glutamic acid | d |
| 2.7 | 25 | 70 | 120 | $NH_4OH$ | e |
| 1.8 | 760 | 50 | 5060 | 1,2 dodecandiol | f |
| 2.5 | 150 | 32 | 1560 | PEG | g |
| 4.6 | 120 | 110 | 364 | PVP | h |
| 3.0 | 140 | 90 | 520 | - | i |
| 2.0 | 84(46) | 120(38) | 233 | PEG | this work† |

a) A Sundaresan and C. N. R. Rao, Nano Today **4** 96 (2009)
b) Y. Liu et al, J. Phys. Cond. Matter, **20** 165201 (2008)
c) M. Y. Ge et al, Appl. Phys. Lett, **93** 062505 (2008)
d) X. Chen et al, Nanotechnology, **20** 115606 (2009)
e) M. Li et al, Appl. Phys. Lett, **94** 112511 (2009)
f) S. Y. Chen et al, J. Phys. Chem C **114** 19576 (2010)
g) K. Ackland et al, IEEE Trans Magn. **47** 3509 (2011)
h) S. Phokha et al, Nanoscale Res. Lett. **7** 425 (2012)
i) N. Paunovic et al, Nanoscale **4** 5469 (2012)

*Calculated using $\mathcal{N}_{eff} = 1/3$; †Standard deviations for 16 samples.

*Figures*

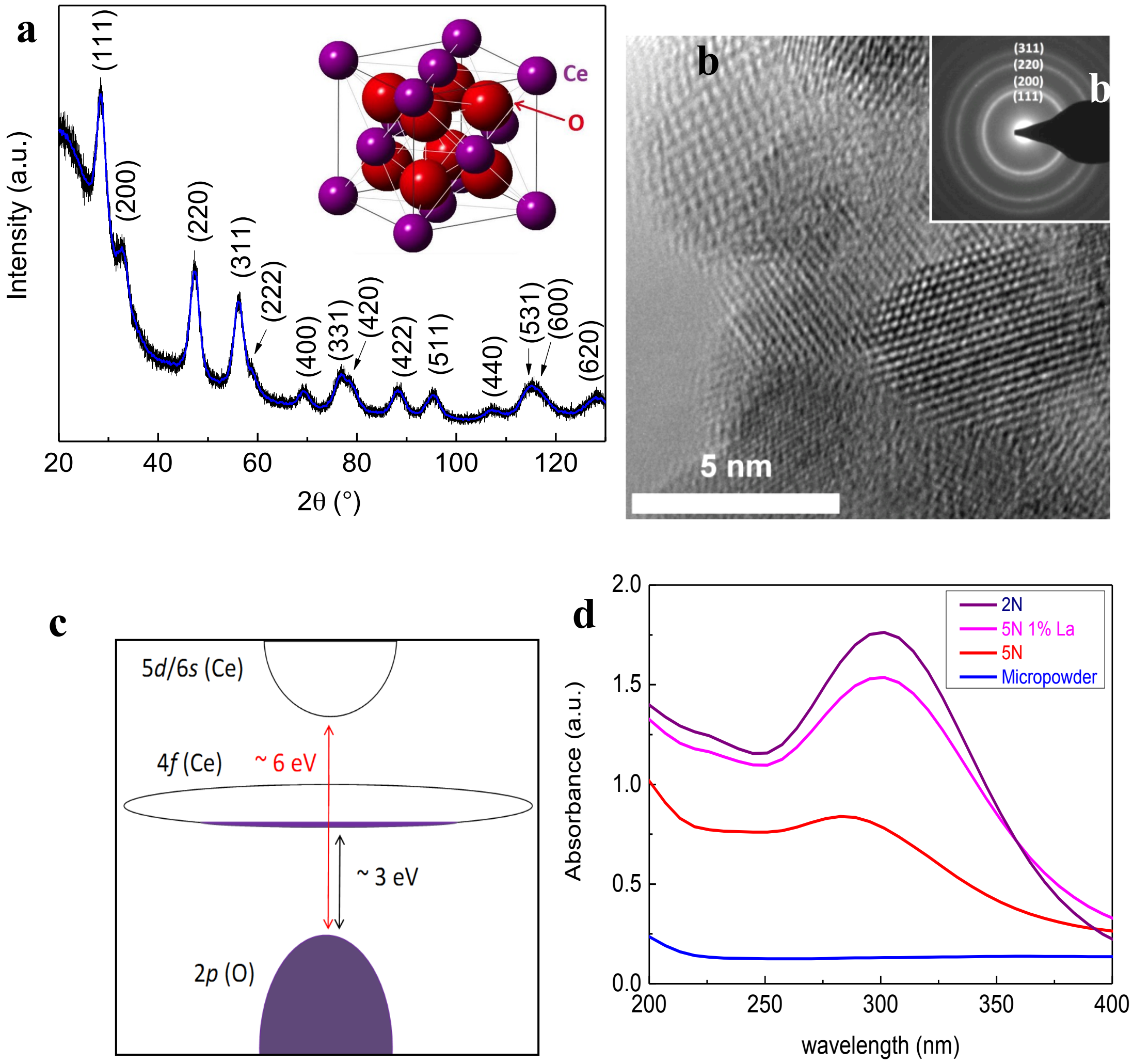

Figure 1. Structure and optical properties of 4 nm $CeO_2$ nanoparticles. a) X-ray diffraction pattern showing the reflections of the fluorite structure, with corresponding particle size broadening, b) Transmission electron micrograph of a few of the particles, c) schematic electronic structure of oxygen deficient $CeO_2$, d) UV-visible absorption spectra of $CeO_2$ micropowders and nanopowders dispersed in 0.1 M $H_2SO_4$. The spectra of the magnetic nanoparticles show a peak at 4 eV.

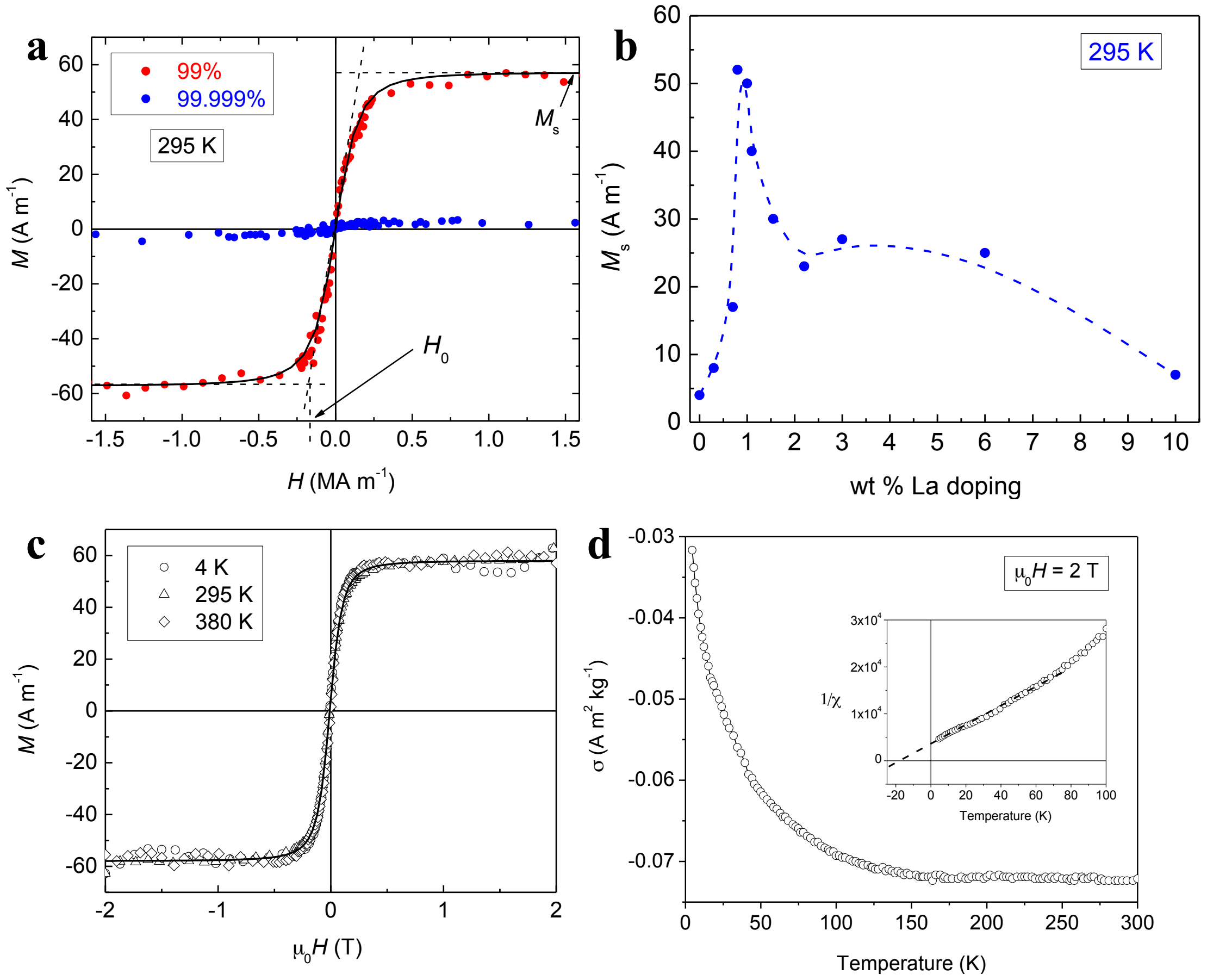

Figure 2. Magnetic properties of 4 mg samples of 4 nm $CeO_2$ nanoparticles. a) room-temperature magnetization curves of samples prepared from 99.999% and 99% precursors; the saturation magnetization $M_s$ and the saturation field $H_0$ are defined as shown, b) The variation of $M_s$ for nanoparticles produced from the pure precursor with La nitrate addition; magnetization is turned on by La substitution, and it is greatest for a La content of 1 %.  c) Magnetization curves of a 99% sample at 4 K, 290 K and 380 K. The data are corrected for the diamagnetism of the sample holder and the high-field susceptibility of the sample; the magnetization curves are fitted to Equation 3, yielding identical parameters at all temperatures. d) The Curie-law

susceptibility deduced from the high-field slope. The plot of susceptibility versus inverse temperature in the insert corresponds to just 0.4 % of $Ce^{3+}$ and a paramagnetic Curie temperature of -8 K, assuming a localized moment of $p_{eff} = 2.54\ \mu_B$ .

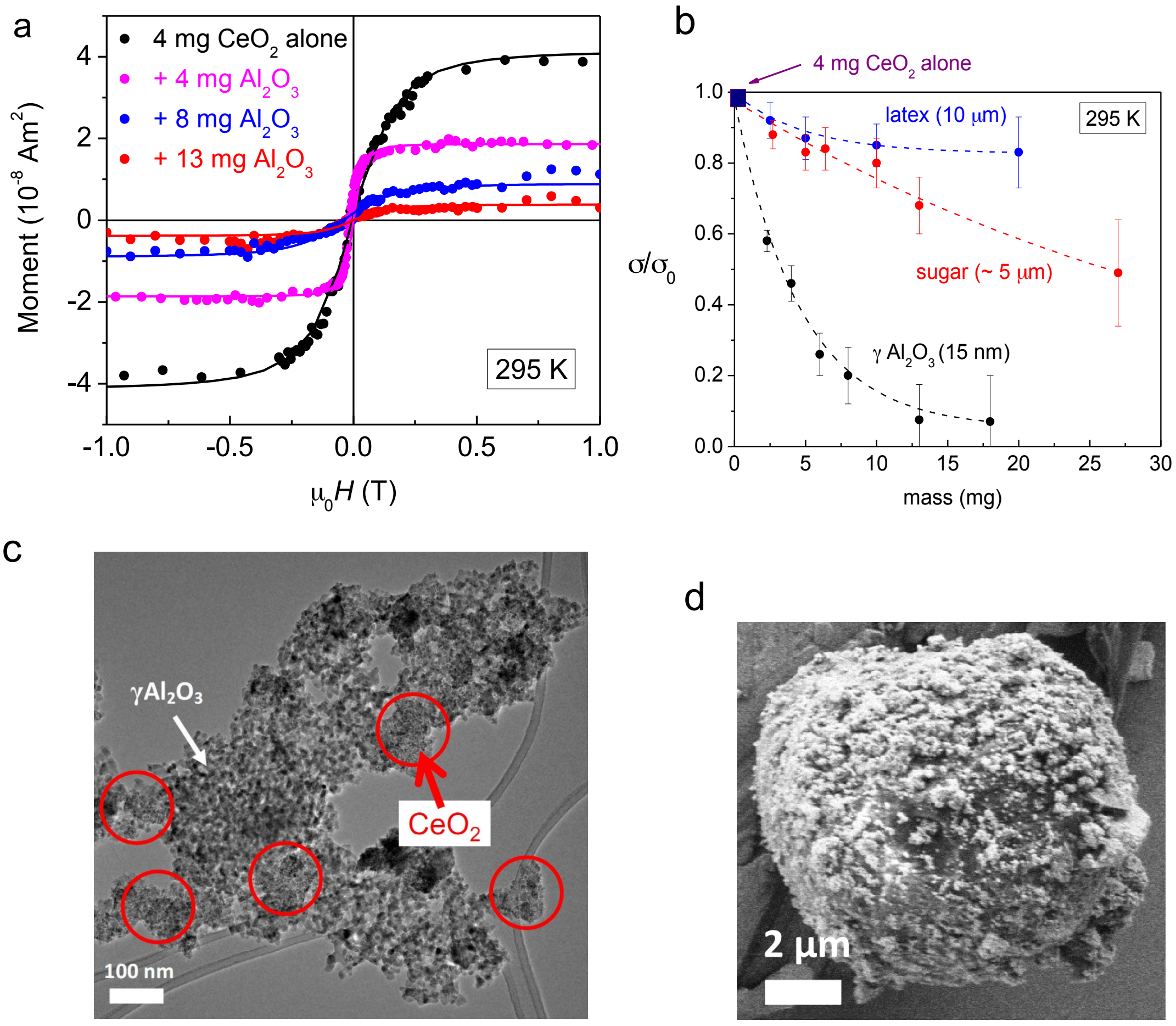

Figure 3. Effect on magnetic moment of diluting the $CeO_2$ with another powder: a) Magnetization curves for a 4 mg sample of 4 nm $CeO_2$ nanopowder diluted with 15 nm $\gamma Al_2O_3$, b) relative magnetization as a function of dilution of 4 mg of 4 nm $CeO_2$ by weight with 15 nm $\gamma Al_2O_3$, 1 μm sugar or 10 μm latex microspheres c) electron micrograph, showing how the dilution with $\gamma Al_2O_3$ breaks the $CeO_2$ down into ~ 100 nm clumps, which destroys the moment and d) the coating of a 10 μm latex microsphere by $CeO_2$ nanoparticles.

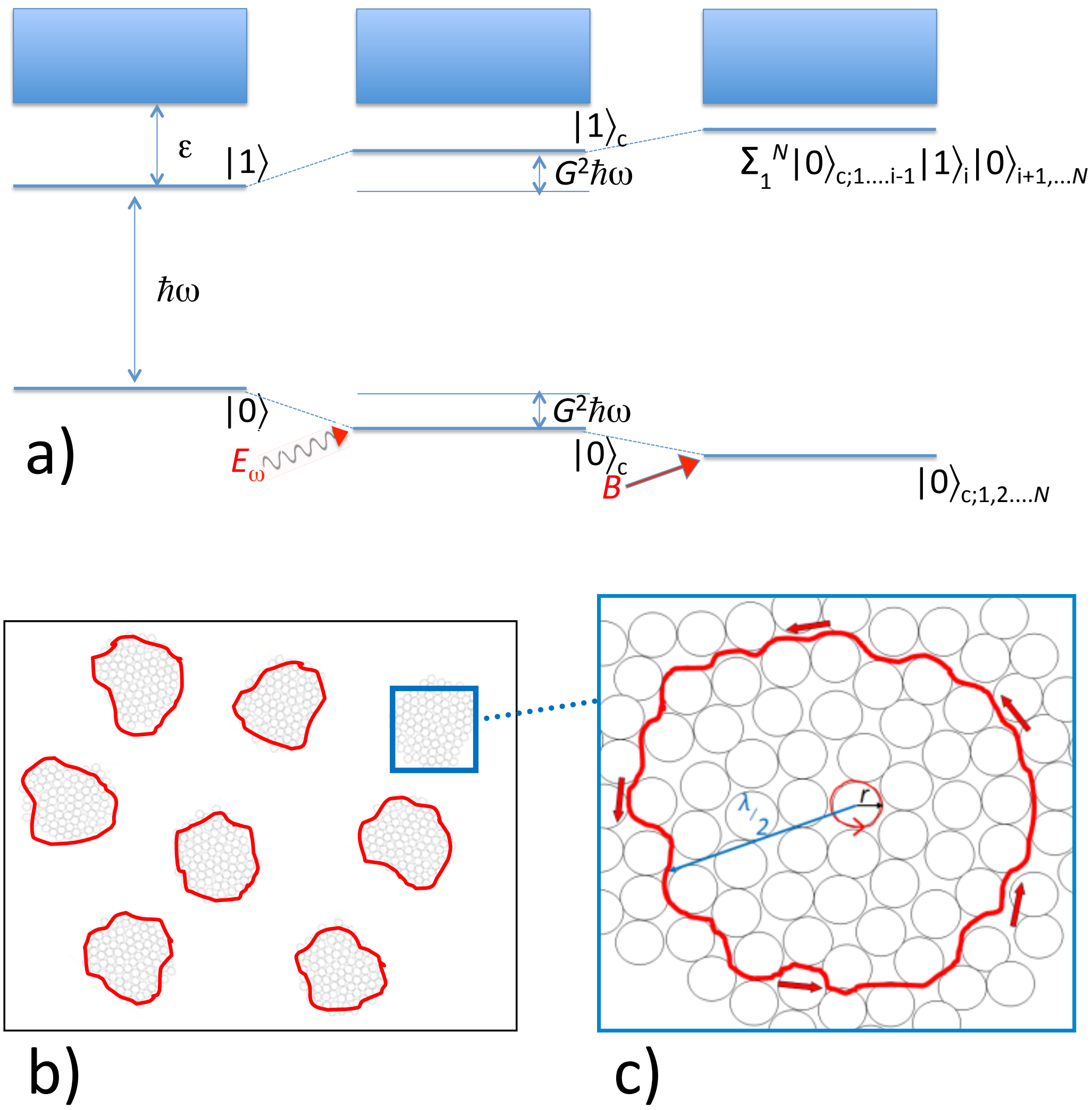

Figure 4. a) Single atom excitations related to impurities or defects in $CeO_2$, showing the effect of coherent interaction with the vacuum electromagnetic field and the consequent influence of a magnetic field on the ground and excited states; b) A sample of $CeO_2$ where a fraction $f$ of the volume is occupied by coherent domains of dimension $\lambda = 2\pi c/\omega$, shaded grey, and c) an orbital current in a coherent domain formed from a clump of $CeO_2$ nanoparticles.

## Supplementary Information

### 1. Orbital moment of the atoms in a coherent domain.

In the theory of Ref. 27, the atoms in a quasi-two-dimensional coherent domain respond collectively to zero-point fluctuations of the vacuum electromagnetic field. The atoms in the model are assumed to have a two-level structure with a baseline ground state $|0\rangle$ at energy 0 and an excited state $|1\rangle$ at energy $\hbar\omega$, which lies $\varepsilon$ below the ionization threshold (See Fig 4a). The effective zero-point electromagnetic field acting an assembly of $N$ two-level atoms in a surface layer is proportional to $\sqrt{N}$, and its effect is to induce collective fluctuations of the electrons at a frequency $G^2\omega$, where $G$ must be of order 0.1 for effects to be observable at room temperature. This leads to small changes of $\pm\, G^2\hbar\omega$ of the ground and excited energy levels, as shown in the figure. The new wavefunctions for an electron in the collective ground state and excited state are

$$|0\rangle_c = |0\rangle \cos\alpha + |1\rangle \sin\alpha \qquad \text{(S1)}$$
$$|1\rangle_c = |1\rangle \cos\alpha - |0\rangle \sin\alpha$$

where the mixing angle $\alpha$ is $\tan^{-1}G$ [27].

We identify the orbital moment of an atom $\mu_c$ as $\frac{1}{2}eG^2\omega r^2$ by analogy with the Bohr atom where the frequency $\omega_B = \hbar/mr_B^2$ and the radius $r_B$ correspond to an orbital moment of $\frac{1}{2}e\hbar/m = \mu_B$, known as the Bohr magneton. We may set $\mu_c = g_c\mu_B$, where $g_c$ is an appropriate g-factor. The electron spin is neglected. The radius $r$ in the expression for $\mu_c$ is related to the ionization energy $\varepsilon$ via the uncertainty principle; $r \approx \hbar/\sqrt{(2m\varepsilon)}$ where m is the electron mass. Hence

$$\mu_c \approx [(2l + 3)(2l + 4)/8]\; eG^2\omega\hbar^2/4m\varepsilon$$
$$= [(2l + 3)(2l + 4)/8]\; (G^2\hbar\omega/2\varepsilon)\mu_B. \qquad \text{(S2)}$$

The numerical factors come from the radial integrals of the wave functions $|0\rangle$ and $|1\rangle$ involved in evaluating $r^2$ in the collective ground state (S1), where $|0\rangle = C_0Y_{l.0}r^l\exp(-r/r_0)\exp(-iG^2\omega t)$ and $|1\rangle = C_1Y_{l+1.0}r^l \exp(-r/r_0)\exp(-iG^2\omega t)$; $Y_{l.m}$ are spherical harmonics, $C_v$ are normalization constants and $r_0 = \hbar/\sqrt{(2m\varepsilon)}$.

In the absence of any external magnetic field, the oscillations are isotropic and there is no net orbital moment.

2. **Induced orbital moment in an applied magnetic field.**

The coupling of the orbital moment of a coherent domain with a static magnetic field $\boldsymbol{B}$ is $-N\boldsymbol{\mu}_c.\boldsymbol{B} = -N\mu_c B \cos\theta = V_c$, where $\theta$ is the average angle between $\boldsymbol{\mu}_c$ and $\boldsymbol{B}$ ($\pi/2$ for $B = 0$). It leads to mixing of the ground state $|0\rangle_c$ and the excited state $|1\rangle_c$ that have been created by interaction with the zero-point field (Eq. S1). The mixing is time-independent since both states have the same time-dependence $\exp(iG^2\omega t)$. The wave function for an electron in the collective magnetically mixed ground state is

$$|0\rangle_{cm} = |0\rangle_c \cos\alpha_m + |1\rangle_c \sin\alpha_m \quad \text{(S3)}$$

where the magnetic mixing angle $\alpha_m$ is $\tan^{-1}(V_c/\lambda_-)$, where $\lambda_- = (\omega/2)[1 - (1 + 4V_c^2/\omega^2)^{1/2}]$ is the energy eigenvalue for the mixed collective ground state, obtained by diagonalizing the effective matrix in the two-state description

$$\begin{vmatrix} -G^2\hbar\omega & V_c \\ V_c & (1+G^2)\hbar\omega \end{vmatrix} \quad \text{(S4)}$$

We have made the approximation $G^2 \approx 0$.

The value of the magnetic interaction with the induced moment in a coherent domain, $\langle 0|-N\mu_c B\cos\theta|1\rangle_c$ includes a factor of the form $2\sin\alpha_m\cos\alpha_m = V_c\lambda_-/(V_c{}^2 + \lambda_-{}^2)$, which reduces to $x/(1 + x^2)^{1/2}$ with $x = 2V_c/\hbar\omega$. The factor $N$ is due to the presence of $N$ mutiparticle states, all of energy $(1 + G^2)\hbar\omega$, which represents a situation where only one elctron is in its excited state and $N$ - 1 are in the coherent ground state (Fig 4a). The average of $\cos\theta$ is $[\kappa G/(1 + G^2)^{1/2}]$, where the factor $\kappa = (l + 1)/[(2l + 1)(2l + 3)]^{1/2}$ comes from evaluating the angular integrals. Hence $x = (2N\mu_c B\cos\theta/\hbar\omega) = (2N\mu_c/\hbar\omega)\,[\kappa G/(1 + G^2)^{1/2}]B$. Since $\kappa \approx \frac{1}{2}$, $x \approx NG\mu_c B/\hbar\omega$, which is the expression used in the main text.

The magnetic moment of the coherent domain is therefore $GN\mu_c x/(1 + x^2)^{1/2}$, which gives Eq. 2 in the main text, with $M_s = G f_c N\mu_c/v_c$, where $f_c$ is the volume fraction of the sample composed of coherent domains, and $v_c$ is the volume of a coherent domain.